\documentclass[aps,showpacs,twocolumn]{revtex4}

\usepackage{bm}
\usepackage{xcolor}
\usepackage{amssymb}
\usepackage{graphicx}

\newcommand{\PRL}[3]{Phys.\ Rev.\ Lett.\ {\bf #1},\ #2 (#3)}

\newcommand{\PRD}[3]{Phys.\ Rev.\ D\ {\bf #1},\ #2 (#3)}
\newcommand{\be}{\begin{equation}}
\newcommand{\ee}{\end{equation}}

\begin{document}

\title{ Pairing in spin polarized two-species fermionic mixtures with mass
asymmetry}

\author{ S. A. Silotri}
\email{silotri@prl.res.in}
\author{ D. Angom}
\email{angom@prl.res.in}
\author{H. Mishra}
\email{hm@prl.res.in}
\affiliation{Theoretical Physics Division, Physical Research Laboratory,
Navrangpura, Ahmedabad 380 009, India}
\author{Amruta Mishra}
\email{amruta@physics.iitd.ac.in}
\affiliation{Department of Physics, Indian Institute of Technology, New
Delhi-110016,India}
\begin{abstract}

We discuss on the pairing mechanism of fermions with mismatch in 
their fermi momenta due to a mass asymmetry. Using a variational
ansatz for the ground state we also discuss the BCS -BEC crossover of
this system.  It is shown that the breached pairing solution with a
single fermi surface is stable in the BEC regime. We also include the 
temperatures effect on the fermion pairing within an approximation that
 is valid for temperatures much below the critical temperature.

\end{abstract}

\maketitle

%%%%%%%%%%%%%%%%%%%%%%%%%%%%%%%%%%%%%%%%%%%%%%%%%%%%%%%%%%%%%%%%%%%%%%%%%%%
%%%%                    Section I: Introduction                      %%%%%
%%%%%%%%%%%%%%%%%%%%%%%%%%%%%%%%%%%%%%%%%%%%%%%%%%%%%%%%%%%%%%%%%%%%%%%%%%%

\def\zbf#1{{\bf {#1}}}
\section{Introduction}

In recent times, pairing in degenerate fermi gas of atoms
\cite{demarco} 
has attracted lot of attention. This is the outcome of the rapid advancement in the 
experimental techniques to study and manipulate systems of ultracold 
atoms. With these techniques, it is possible to cool and trap one or more 
hyperfine states of an element and control the population in each of these 
states. Furthermore, the interaction strength
as well as the sign of the interaction
between two components can be tuned over a wide range, using the techniques of
Feshbach resonance \cite{regal}. When the coupling is weak and attractive, 
the fermion system can be successfully described with the Bardeen-Cooper-
Schrieffer (BCS) theory. The clinching evidences are the experimental
measurement of the gap energy \cite{chin} and observations of vortices
\cite{zwierlein}. Typically, in such situations the coherence length 
is much larger than the interparticle separation. However, the picture 
changes as the coupling strength is increased. The Cooper pairs are more 
localized and the superfluidity is realized by Bose Einstein condensation 
(BEC) of molecular boson comprising of a pair of fermions. It is further  
expected that such a phenomenon is a cross over between the BCS and BEC 
regimes. These studies have been extended to include two fermion species
with imbalanced populations. In such cases, instead of a crossover, the 
system is expected to show a very interesting and rich phase structure with 
the appearance of exotic superfluids. These include the existence of interior 
gap superfluidity with one fermi surface, breached pairing with two fermi 
surfaces. It is also possible to have inhomogeneous phases like 
Larkin-Ovchhinnikov-Fulde-Ferrel (LOFF) phase wherein the Cooper
pairs have nonzero net momentum \cite{nardulli}, superfluidity with 
deformed Fermi surfaces \cite{sedrakian}
or a phase separated state \cite{caldas}. These exotic phases 
emerge when pairing occurs between two species whose Fermi surfaces do not
match. This can happen when the number densities of the two 
species are different or there is a mismatch in their masses or both 
\cite{wilczek,shin,sheehy,parisha,yip,parishb}. Till date, the two fermion 
species experiments are with 
the two hyperfine states of the same alkali atom forming the condensate 
\cite{ketterle}. Very recently two fermion species of different masses,
lithium and potassium, were laser cooled and trapped to degeneracy
\cite{taglieber}. And another recent work reports the observations of 
Feshbach resonances \cite{wille} with the same system. Thus 
achievement of superfluidity with this mass difference could be the next 
frontier of ongoing experiments in ultracold fermions.

In this paper, we attempt to describe such a system by constructing a 
variational ground state explicitly and the gap function in the analysis is 
determined by minimization of the thermodynamic potential with the constraints 
of fixed particle number densities for the two species. Minimization of the 
thermodynamic potential decides which phase is preferred at the given 
densities of pairing species.
This method has earlier been considered to describe cold 
fermionic atoms with equal masses for homogeneous \cite{rapid} as well as
inhomogeneous pairing \cite{amhmloff}. This method has also been applied
to relativistic system like cold quark matter and color superconductivity
\cite{amhmq}.
In the present work, we discuss the possible structures with
homogeneous pairing both with density asymmetry as well as mass asymmetry
for the two condensing species.

We organize the present work as follows. In section \ref{ansatz}
we discuss the ansatz for the ground state and the  Hamiltonian 
in terms of the Four fermi point interaction to model
the superfluidity for the two species of fermionic atoms. 
In section \ref{eval}
we evaluate the thermodynamic potential by minimizing the thermodynamic
potential with respect to the functions in the ansatz for the
``ground state". In section \ref{results} we discuss the results regarding 
asymmetric fermionic populations and the gapless phases. Finally we 
summarize and conclude our results in section \ref{summary} .

%%%%%%%%%%%%%%%%%%%%%%%%%%%%%%%%%%%%%%%%%%%%%%%%%%%%%%%%%%%%%%%%%%%%%%%%%%%
%%%%      Section II: Ansatz for the ground state and the hamiltonian    %%
%%%%%%%%%%%%%%%%%%%%%%%%%%%%%%%%%%%%%%%%%%%%%%%%%%%%%%%%%%%%%%%%%%%%%%%%%%%

\section{Ansatz for the ground state and the Hamiltonian
\label{ansatz} }
To examine the superfluidity for fermionic atoms, we 
consider a Hamiltonian describing two interacting fermionic
species with four-fermion point interaction  given as
\begin{eqnarray}
    H & = & \sum_{i}\Psi_{r}^{i^{\dagger}}(\mathbf{z})\left(-
       \frac{\hbar^{2}{\vec \nabla}^{2}}{2m_i}\right)\Psi_{r}^{i}(\mathbf{z})
                          \nonumber \\
       & + &  \sum_{r,s}g\Psi_{r}^{1^{\dagger}}(\mathbf{z})\Psi_{r}^{1}
            (\mathbf{z})\Psi_{s}^{2^{\dagger}}(\mathbf{z})\Psi_{s}^{2}
            (\mathbf{z}),
  \label{ham}
\end{eqnarray}
where $r$ and $s$ are the spin indices and $i$ denotes 
the species with mass $m_i$. 
The constant $g$ is the bare interaction strength between the two species
and is related to the $s$-wave scattering length $a$. To describe  pairing
between two different fermionic species, we consider the ansatz for the
ground state \cite{rapid,amhmloff} of the system as
\begin{equation}
  |\Omega\rangle=e^{(B^{\dagger}-B)}|0\rangle ,
  \label{eq:}
\end{equation}
 where $B^{\dagger}=\int d\mathbf{k}\epsilon^{ij}\Psi_r^{i^{\dagger}}
(\mathbf{k})f(k)\Psi_{-r}^{j^{\dagger}}(-\mathbf{k})$.
Here $\epsilon^{ij}$ is the Levi-Cevita tensor, with $i$ and $j$
denoting two different fermionic species. The function $f(k)$ is the 
variational function related to the order parameter, as will be seen later. 
In the case of equal population and for negative weak coupling, this ansatz 
corresponds to the standard BCS wave function. The two ground states, $|0\rangle$ and 
$|\Omega\rangle$, are related by the unitary transformation operator
$U=e^{(B^{\dagger}-B)}.$ Hence the field operators transform as 
$\Psi'=U\Psi U^{\dagger}$ where as $\Psi '$ is the annihilation operator
for $|\Omega\rangle$. To include the effect of temperature and density,
we use the method of thermo-field dynamics (TFD) that is particularly useful while 
dealing with operators and states. Here, the thermal ``ground state" is
obtained from $|\Omega\rangle$ through a Bogoliubov transformation in an 
extended Hilbert space associated with thermal doubling of operators. 
Explicitly, $|\Omega,\beta,\mu\rangle$, the ground state at finite temperature 
and density is given as
\be
|\Omega,\beta,\mu\rangle=\exp(B_{\beta,\mu}^\dagger-B_{\beta,\mu})|\Omega\rangle
\label{gs}
\ee
where,
\be
B_{\beta,\mu}^\dagger=\int\left[\Psi'(\zbf k)^\dagger
\theta_-^i(\zbf k,\beta,\mu)\underline{\Psi'}(-\zbf k)\right]d\zbf k .
\label{bbta}
\ee
In Eq.(\ref{bbta}), the function $\theta^i$, as we shall see later,
will be related to the
distribution function of the $i^{\rm{th}}$ species and the underlined operators are 
the operators in the extended Hilbert space associated with thermal doubling.
All the functions in the ansatz in Eq.(\ref{gs}), the condensate function
$f(\zbf k)$, the thermal functions $\theta^i(\zbf k,\beta,\mu)$ shall be determined by 
extremizing the thermodynamic potential. We carry out this 
extremization in the next section.

%%%%%%%%%%%%%%%%%%%%%%%%%%%%%%%%%%%%%%%%%%%%%%%%%%%%%%%%%%%%%%%%%%%%%%%%%%%%%%%%
%%%%  Section III: Evaluation of Thermodynamic potential and the gap equation %%
%%%%%%%%%%%%%%%%%%%%%%%%%%%%%%%%%%%%%%%%%%%%%%%%%%%%%%%%%%%%%%%%%%%%%%%%%%%%%%%%

\section{Evaluation of Thermodynamic potential and the gap equation
\label{eval}}
Having defined the ground state as in eq.(\ref{gs}), we next evaluate the
thermodynamic potential corresponding to the Hamiltonian
given in Eq.(\ref{ham}). To calculate, e.g., the energy density, one can take 
the expectation value of the Hamiltonian. %Using the fact that, 
Noting that the variational
state in Eq.(\ref{gs}) arises from successive Bogoliubov transformations,
one can calculate the expectation values of the various operators. Thus we have,
with  $\langle \hat O\rangle$ representing
the expectation value of an operator $\hat O$ in the new ground state of the
system $\langle\Omega,\beta,\mu)|\hat O|\Omega,\beta,\mu\rangle$,

\begin{eqnarray}
   \left\langle \Psi_r^{1\dagger}(\mathbf{k}_{1})\Psi_s^1(\mathbf{k}_2)
   \right\rangle & = & \left[\cos^2(f(k_1))\sin^2(\theta_1(k_1))+
              \sin^2(f(k_1))\right. \nonumber \\
 &  & \left . \cos^2(\theta_2(k_1))\right]\delta_{rs}\delta\left(\mathbf{k}_{1}
     -\mathbf{k}_{2}\right)\label{expect1}\\
   \left\langle \Psi_r^{2\dagger}(\mathbf{k}_1)\Psi_s^2(\mathbf{k}_{2})
   \right\rangle & = & \left[\cos^2(f(k_1))\sin^2(\theta_2(k_1))
    -\sin^{2}(f(k_{1})) \right. \nonumber \\
  &&  \left.\cos^2(\theta_1(k_1))\right]\delta_{rs}\delta(\bm{k}_1-\bm{k}_2),
     \label{expect2}  \\
    \left\langle \Psi_r^1(\mathbf{k}_{1})\Psi_s^2(\mathbf{k}_{2})
    \right\rangle  & = & -\frac{\sin(2f(k_1))}{2}[1-\sin^2(\theta_1(k_1))
      \nonumber \\
  &-&  \sin^2(\theta_2(k_1))]\delta_{r-s}\delta(\bm{k}_1+\bm{k}_2), 
       \label{expect3}\\
    \left\langle \Psi_r^{1\dagger}(\mathbf{k}_1)\Psi_s^{2\dagger}
    (\mathbf{k}_2)\right\rangle  & = & \frac{\sin(2f(k_1))}{2}\left[1-
    \sin^2(\theta_1(k_1))  \right .
           \nonumber \\ 
  &-& \left. \sin^2(\theta_2(k_1))\right]\delta_{r-s}\delta(\bm{k}_1+
      \bm{k}_2).
    \label{expect4}
\end{eqnarray}
Note that the thermodynamic potential is given by
\begin{equation}
   \Omega=\epsilon-\frac{s}{\beta}-\mu_i\rho^i ,\label{omg}
\end{equation}
where, $\epsilon=\langle H\rangle_{\beta,\mu}=T+V$ is the energy density, with $T$ and
$V$ as the kinetic and potential energy contributions respectively, $s$ is the entropy density
and $\mu_i$ is the chemical potential for the species $`i`$.
The diagonal part of the potential in Eq. (\ref{omg}) is given by
\begin{eqnarray}
   T-\mu N &=&
\sum_i\Psi^{i\dagger}(\mathbf{z})(\varepsilon_i-\mu_i)\Psi^i(
           \mathbf{z})\nonumber\\
& =&
   \frac{1}{(2\pi)^3}\int d^3k \left[\xi'_1[\cos^2(f)\sin^2(\theta_1)\right.\nonumber \\
  &+&\sin^2(f)\cos^2(\theta_2)]+ \xi'_2\left[\cos^2(f)\sin^2(\theta_{2})
       \right.\nonumber \\
  &+& \left.\left.\sin^{2}(f)\cos^{2}(\theta_{1})\right]\right],
  \label{T}
 \end{eqnarray}
where, $\xi_{i}^{'}=\hbar^2k^2/2m_i-\mu_i$ is the kinetic energy 
with respect to the chemical potential,
of the $i^{\rm th}$ species.
Similarly the expectation of the
term $H_{\rm int}$ is simplified using the Wick's theorem 
\begin{equation}
   V\equiv\langle H_{\rm int}\rangle=g\rho_{1}\rho_{2}+gI_{D}^{2}, 
  \label{V}
\end{equation}
$I_{D}$ is related to the condensate defined as 
$I_D=\delta_{r-s}\left\langle \Psi_r^{1\dagger}(\mathbf{k})
\Psi_s^{2\dagger}(\mathbf{-k})\right\rangle$ and is given as 
\be
I_D=\frac{1}{(2\pi)^3}\int d\zbf k \sin 2 f(\zbf k)\left(\cos^2\theta_1-\sin^2\theta_2\right).
\label{id}
\ee
Further, the species densities $\rho_i=\left\langle 
\Psi_r^{i\dagger}(\mathbf{k})\Psi_s^i(\mathbf{k})\right\rangle\delta_{rs}$
for the fermions are  given as
\begin{eqnarray}
   \rho_1 &= &\frac{1}{(2\pi)^3}\int d^3k\left[\cos^2(f)\sin^2(
               \theta_1) \right. \nonumber \\
   && \left. +\sin^2(f)\cos^2(\theta_2)\right],\label{rho1} \\
   \rho_2 & = &\frac{1}{(2\pi)^3}\int d^3k\left[\cos^2(f)\sin^2(\theta_2)
               \right.\nonumber \\
   &  & \left.+\sin^2(f)\cos^2(\theta_{1})\right].\label{rho2}
\end{eqnarray}
Finally the entropy density for the two species fermionic mixture is \cite{tfd} 
\begin{eqnarray}
   s & = & -\sum_{i=1,2}\frac{1}{(2\pi)^3}\int d^3k\left[n_i(k)\log(n_i(k))+
            \right.\nonumber \\
      &  & \left.(1-n_i(k))\log(1-n_i(k))\right],
  \label{expects}
\end{eqnarray}
where $n_i(k)=\sin^2(\theta_i)$ is the density distribution of the 
$i^{\rm{th}}$ species. Combining Eqs.(\ref{T}), (\ref{V}), and 
(\ref{expects}), one can then calculate the expectation value of the
thermodynamic potential in the ansatz state given in Eq.(\ref{gs}).
The thermodynamic potential is a functional of three functions,
the condensate function, $f(k)$ and the two thermal distribution 
functions $\theta_{i}(k)$ for the two species. These functions 
are determined by functional minimization of the thermodynamic 
potential of Eq.(\ref{omg}) which we shall analyse in the next subsection.

%%%%%%%%%%%%%%%%%%%%%%%%%%%%%%%%%%%%%%%%%%%%%%%%%%%
%%%%  Section A: Gap equation                %%%% 
%%%%%%%%%%%%%%%%%%%%%%%%%%%%%%%%%%%%%%%%%%%%%%%%%%%

\subsection{Gap equation}

 Functional minimization of the thermodynamic potential $\Omega$ with 
respect to $f(k)$ gives 
\begin{equation}
  \tan(2f(k))=-\frac{2gI_D}{(\varepsilon_1+\varepsilon_2)
              -(\nu_1+\nu_2)}
\equiv\frac{\Delta}{\bar\epsilon-\bar\nu},
  \label{tan2fk}
\end{equation}
where $\nu_i=\mu_i-g\epsilon^{ij} \left|\rho_j\right|$
is the chemical potential with the mean field correction. 
We have also defined in the above, the superconducting gap
$\Delta=-gI_D$ and $\bar\epsilon=
(\epsilon_1+\epsilon_2)/2$, $\bar\nu=(\nu_1-\nu_2)/2$ as the average
kinetic energy and chemical potential respectively. Let us note that,
the condensate function depends on the {\em average kinetic energy} and
the {\em average chemical potentials} of the two condensing species.
Substituting the solution for the condensate function from Eq.(\ref{tan2fk}) in
the definition of $I_{D}$ given by Eq.(\ref{id}), 
we obtain the gap equation

\begin{eqnarray}
  \Delta  =  -\frac{g}{(2\pi)^3}\int d^3 k\frac{\Delta}
                {2\omega}\left[1-\sin^2(\theta_1)-
          \sin^2(\theta_2)\right].
  \label{gap1}
\end{eqnarray}
Similarly, one obtains the densities for the two fermion species as
\begin{eqnarray}
  \rho_{1} & = & \frac{1}{(2\pi)^{3}}\int d^3k\left[\frac{1}{2}
                 \left(1+\frac{\xi}{\omega}\right)\sin^2(\theta_1)\right.
		 \nonumber \\
           & + & \left.\frac{1}{2}\left(1-\frac{\xi}{\omega}\right)\cos^{2}
	        (\theta_{2})\right], \\
  \rho_{2} & = & \frac{1}{(2\pi)^{3}}\int d^3k\left[\frac{1}{2}
                  \left(1+\frac{\xi}{\omega}\right)\sin^2(\theta_2)\right.
		  \nonumber \\
            & + & \left.\frac{1}{2}\left(1-\frac{\xi}{\omega}\right)\cos^2
	         (\theta_1)\right],
\end{eqnarray}
where, we have denoted $\xi=\epsilon-\nu$.
The other variational extremisation conditions  
$\delta\langle\Omega\rangle/\delta\theta_i(k)=0$ determine 
 $\theta_{i}(k)$ to be related to the distribution functions
for the ith fermion species as
\begin{equation}
  \sin^2(\theta_i(k,\mu))=\frac{1}{\exp(\beta\omega_i)+1}.
\end{equation}
In the above, the quasiparticle energies 
are given as 
$\omega_1=\omega+\delta_\xi$ and  $\omega_2=\omega-\delta_\xi$ 
where $\delta_\xi=[(\varepsilon_1-\nu_1)-(\varepsilon_2-\nu_2)]/2$.
 
Next we examine the gap equation Eq.(\ref{gap1}), which for  
nonzero $\delta$  can be written as
\begin{equation}
  -\frac{1}{g}=\frac{1}{(2\pi)^3}\int d^3k\frac{1}{2\omega}
               \left[1-\sin^2(\theta_1)-\sin^2(\theta_2)\right].
  \label{g}
\end{equation}
This equation is ultraviolet divergent which is characteristic of the 
contact interaction.
It is rectified by subtracting the contribution at $T=0$ and $\mu=0$ 
and relating this renormalized coupling to the s-wave scattering length $a$ 
\cite{randeria,rapid}.
Thus regularized gap equation is 
\begin{eqnarray}
  -\frac{\tilde{m}}{4\pi\hbar^{2}a}
    & = & \frac{1}{(2\pi)^3}\int d^3k\left
        (\frac{1}{2\omega}\left[1-\sin^2(\theta_1)
	\right.\right.\nonumber \\
     & - & \left.\left.\sin^2(\theta_2)\right]-\frac{1}{2\varepsilon_k}\right),
  \label{gap}
    \end{eqnarray}
    where $\tilde{m}$ is the reduced mass. 

%%%%%%%%%%%%%%%%%%%%%%%%%%%%%%%%%%%%%%%%%%%%%%%%%%%
%%%%  Section IIc: Stability condition         %%%% 
%%%%%%%%%%%%%%%%%%%%%%%%%%%%%%%%%%%%%%%%%%%%%%%%%%%

\subsection{Stability condition}

The stability of the pairing state is decided by comparing the thermodynamic
potential of the superconducting matter with that of the normal matter. Thus the 
relevant quantity is the difference of the thermodynamic potential between the paired and
normal phases. The thermodynamic potential of the paired fermionic mixture 
is
\begin{eqnarray}
  \Omega & = & \frac{1}{(2\pi)^3}\int d^3k\left[\xi-\omega-\frac{1}{\beta}
               \sum_i \ln\left(1+\exp\left(-\beta\omega_i
               \right)\right)\right]\nonumber. \\
   &  & -\frac{\Delta^2}{g}-g\rho_1\rho_2.
 \end{eqnarray}
Subtracting the  thermodynamic potential for normal matter 
$\left(\Delta=0\right)$
from the above equation, we have the difference in the thermodynamic potential
between the condensed and the normal matter as
\begin{eqnarray}
  \delta\Omega & = & \frac{1}{(2\pi)^3}\int d^3k\left[\left|\xi\right|-
                     \omega-\frac{1}{\beta}\sum_i\ln\left(1+ \exp\left(-
                     \beta\omega_i\right)\right)\right.\nonumber \\
   &  & \left.+\frac{1}{\beta}\sum_i\ln\left(1+\exp\left(-\beta\omega_{0i}
         \right)\right)\right]-\frac{\Delta^2}{g}.
\label{delomg}
\end{eqnarray}
 Here $\omega_i=\omega\pm\delta_\xi$ and 
$\omega_{0i}=\left|\xi\right|\pm\delta_\xi$.  This difference in the thermodynamic potential,
 $\delta\Omega$ has to be negative for the stability of the paired state.
Further one can use the gap 
equation to eliminate  the coupling $g$ in Eq.(\ref{delomg}) 
to obtain
\begin{eqnarray}
  \delta\Omega & = & \frac{1}{(2\pi)^3}\int d^3k\left[\left|\xi\right|-
                     \omega+ \frac{\Delta^2}{2\omega}-\frac{1}{\beta}
                     \sum_i \ln\left(1+\right.\right.\nonumber \\
                &  & \left.\exp\left(-\beta\omega_i\right)\right)+\frac{1}
	             {\beta}\sum_i\ln\left(1+\exp\left(-\beta\omega_{0i}
		     \right)\right)\nonumber \\
                &  & \left.-\frac{\Delta^2}{2\omega}\left[\sin^2
	            (\theta_1)+\sin^2(\theta_2)\right]\right].
  \label{omega}
\end{eqnarray}
This expression is free of any ultraviolet divergence and will be used 
to determine the stability of the given paired state.

%%%%%%%%%%%%%%%%%%%%%%%%%%%%%%%%%%%%%%%%%%%%%%%%%%%%%%%%%%%%%%%%%%%%%%%%%%%
%%%%            Section III: Zero temperature limit                   %%%%%
%%%%%%%%%%%%%%%%%%%%%%%%%%%%%%%%%%%%%%%%%%%%%%%%%%%%%%%%%%%%%%%%%%%%%%%%%%%
\subsection{Zero temperature limit}

In the limit of zero temperature, the quasi-particle density distribution
of the two atomic species 
\begin{equation}
  n_i(k)=\lim_{\beta\rightarrow\infty}\frac{1}{\exp(\beta\omega_i)+1}=
         \Theta(-\omega_{i}),
\end{equation}
where $\Theta(\ldots)$ is the Heaviside step function. The gap equation
in this limit is given as
\begin{eqnarray}
   -\frac{\tilde{m}}{4\pi\hbar^{2}a}&=&\frac{1}{(2\pi)^3}
          \int d^3k\left [ \frac{1}{2\omega}\big(1-\Theta(-\omega_1)-
          \Theta(-\omega_2) \big) \right. 
               \nonumber \\
      && \left. -\frac{1}{2\varepsilon_{k}}\right].
  \label{gap2}
\end{eqnarray}
The densities of the two atomic species are 
\begin{eqnarray}
   \rho_1 & = & \frac{1}{(2\pi)^3}\int d^3k\left[\frac{1}{2}
                \left(1+\frac{\xi}{\omega}\right)\Theta(-\omega_{1}) \right.
                      \nonumber \\
           &+&   \left. \frac{1}{2}\left(1-\frac{\xi}{\omega}\right)(1-\Theta(-
                \omega_{2}))\right]
               \label{zerorho1}\\
   \rho_2 & = & \frac{1}{(2\pi)^3}\int d^3k\left[\frac{1}{2}
                \left(1+\frac{\xi}{\omega}\right)\Theta(-\omega_{2}) \right.
                      \nonumber \\
           &+&   \left. \frac{1}{2}\left(1-\frac{\xi}{\omega}\right)(1-\Theta(-
                \omega_{1}))\right]. \label{zerorho2}
\end{eqnarray}

For numerical calculations, it is useful to express 
the equations Eqs. (\ref{omega}) and (\ref{gap2})--(\ref{zerorho2}) in terms of dimensionless 
quantities. Hence we make the substitutions 
%$|\zbf k|=k_F x$,
$k=k_F x,$ $\Delta=\epsilon_F \hat{\Delta},$ 
$\nu=\epsilon_F \hat{\nu},$  $\delta_\nu=\epsilon_F \hat{\delta_\nu}$ and
$\omega=\epsilon_F \hat{\omega}$ where
$k_{F}$ is the Fermi momentum defined as 
$k_{F}^3=3\pi^2(\rho_{1}+\rho_{2})$ and 
$\epsilon_{F}=\hbar^2k_{F}^2/2\tilde{m}.$
In terms of the dimensionless quantities
\begin{equation}
   -\frac{\pi}{k_Fa}=\int_0^{\infty}\frac{x^2dx}{\hat\omega}\left[1-\Theta(-
    \hat\omega_1)-\Theta(-\hat\omega_2)-\frac{1}{2}\right],
  \label{scaledgape}
\end{equation}
\begin{eqnarray}
   \frac{\rho_1}{k_F^3} & = & \frac{1}{2\pi^2}\int_0^{\infty}x^{2}dx
         \left[\frac{1}{2}\left(1+\frac{\hat\xi}{\hat\omega}\right)\Theta(-\hat\omega_1)
             \right. \nonumber \\
        && \left. +\frac{1}{2}\left(1-\frac{\hat\xi}{\hat\omega}\right)(1-\Theta(-
           \hat\omega_{2}))\right]   \label{finalrho1}\\
    \frac{\rho_2}{k_F^3} & =&\frac{1}{2\pi^2}\int_{0}^{\infty}x^2dx
         \left[\frac{1}{2}\left(1+\frac{\hat\xi}{\hat\omega}\right)\Theta(-\hat\omega_2)
             \right. \nonumber \\
        && \left. +\frac{1}{2}\left(1-\frac{\hat\xi}{\hat\omega}\right)(1-\Theta(-
           \hat\omega_1))\right].\label{finalrho2}
\end{eqnarray}
The Eqs.(\ref{scaledgape})-(\ref{finalrho2}) are the governing equations
of the paired state for two component fermionic mixture at zero temperature,
when the components are homogeneous and not equal in density. These
three equations are to be solved self consistently. In the zero temperature 
limit, the difference in the thermodynamic potential 
\begin{eqnarray}
   \delta\Omega & = & \frac{1}{(2\pi)^3}
\int d^3k\left[|\xi|-\omega +\frac{\Delta^2}{2\omega}\right]+
         \frac{1}{(2\pi)^3}\int d^3k 
\Bigg [ \Bigg (\omega_1 
              \nonumber \\
     && -\frac{\Delta^2}{2 \omega}\Bigg )\Theta(-\omega_1)
        +\left(\omega_2-\frac{\Delta^2}{2\omega}\right)\Theta(-\omega_2)
        \Bigg] \nonumber \\
     && -\frac{1}{(2\pi)^3}\int d^3k\Bigg [\omega_{01}\Theta(-\omega_{01})+
        \omega_{02}\Theta(-\omega_{02})\Bigg ].
\label{omega_zero}
\end{eqnarray}
In the above, $\omega_1=\omega +\delta_\xi,$ $\omega_2=\omega -\delta_\xi,$ 
$\omega_{01}=\left|\xi\right|+\delta_\xi$ and
$\omega_{02}=\left|\xi\right|-\delta_\xi.$
%$\omega_{0\alpha}=\left|\xi\right|\pm\delta_\xi.$

%%%%%%%%%%%%%%%%%%%%%%%%%%%%%%%%%%%%%%%%%%%%%%%%%%%%%%%%%%%%%%%%%%%%%%%%%%%
%%%%            Section IV: Breached pair solution                   %%%%%
%%%%%%%%%%%%%%%%%%%%%%%%%%%%%%%%%%%%%%%%%%%%%%%%%%%%%%%%%%%%%%%%%%%%%%%%%%%

\subsection{Breached Pair solution}

For the symmetric case of two fermionic species of equal masses, 
having equal densities,  $\delta_\xi$ is zero.
Then, $\omega_{1}=\omega_{2}=\omega$ and the $\Theta(-\omega_{i})$
in the equations are zero as $\omega_{i}\geq 0$ always. This is the
standard BCS phase. In the general case when $\delta_\xi$ is nonzero,
one of the $\Theta(\ldots)$ functions in the equations has a nonzero 
contribution. This can
occur when there is a difference between the densities or a difference
in the masses of
the two fermion species. Without loss of generality, let  us consider the case when
 $\delta_\xi$ is negative.  In this case $\omega_{2}=\omega-\delta_\xi$ is 
always positive and $\Theta\left(-\omega_{2}\right)$ will be zero. However, 
$\omega_{1}(=\omega+\delta_{\xi})$ is negative when $\omega<|\delta_\xi|$ 
and in this domain $\Theta\left(-\omega_{1}\right)$ is nonzero.
The quasi particle excitations become gapless at momenta $k_{\rm max}$ and 
$k_{\rm min}$ given as

\begin{eqnarray} 
   \hbar^{2}k_{\rm max/min}^{2}&=&(m_{1}\nu_{1}+m_{2}\nu_{2})
            \nonumber \\
    &\pm& \sqrt{(m_{1}\nu_{1}-m_{2}\nu_{2})^{2}-4m_{1}m_{2}\Delta^{2}}.
    \label{kminmax}
\end{eqnarray}
The fermionic mixture supports a gapless mode in the momentum domain
$(k_{\rm min}^{2},k_{\rm max}^{2})$, in scaled units 
$x_{\rm min}=k_{\rm min}/k_{F}$ and 
$x_{\rm max}=k_{\rm max}/k_{F}$.

It is convenient to reduce the density equations to the total and the difference of 
the densities. Using these, one can define the experimentally relevant 
parameter polarization $P=(\rho_{1}-\rho_{2})/(\rho_{1}+\rho_{2})$, the 
measure of population difference of the two fermionic species.  Since 
only $\Theta(-\omega_{1})$ is nonzero, the difference of the densities 
\begin{equation}
    \frac{2\pi^2(\rho_1-\rho_2)}{k_F^3}=\int_{x_{\rm min}}^{
       x_{\rm max}}x^2dx=\frac{1}{3}(x_{\rm max}^{3}-x_{
       \rm min}^{3}).
\end{equation}
Rearranging the above equation, we get 
\begin{equation}
   \delta\rho=\rho_1-\rho_2=\frac{k_F^3}{6\pi^2}(x_{\rm max}^3-x_{\rm min}^3).
\label{bpdrho}
\end{equation}
The total density in dimensionless units 
\begin{eqnarray}
   \frac{\rho}{k_F^3}&=&\frac{1}{2\pi^{2}}\int x^2dx\left[\Theta(-
       \hat{\omega}_1)+\left(1- \right.  \right.
               \nonumber \\
      &&  \left. \left. \frac{(x^2/2-\hat{\nu})}{\sqrt{(x^2/2- \hat{\nu})^2+
         \hat{\Delta}^2}}\right)(1-\Theta(-\hat{\omega}_1))\right].
  \label{bpavrho}
\end{eqnarray}
 To solve Eqs.(\ref{scaledgape})--(\ref{bpavrho}), the parameters of the
interacting fermions: scattering length $a$, density of atoms $\rho$
and masses of the atoms $m_{i}$ are to be chosen appropriately. This
can be achieved using the dimensionless quantity $k_{F}a$ with the
definition $\overline{\rho}/k_{F}^{3}=1/3\pi^{2}$. Then the total
density 
\begin{eqnarray}
   \frac{2}{3}&=&\int_0^{\infty}x^2dx\left(1- \frac{x^2/2-\hat{\nu}}{
                 \sqrt{(x^2/2-\hat{\nu})^2+ \hat{\Delta}^2}}\right) 
                   \nonumber \\
        &+&  \int_{x_{\rm min}}^{x_{\rm max}}x^2dx\frac{x^2/2- 
            \hat{\nu}}{\sqrt{ (x^2/2-\hat{\nu})^2+\hat{\Delta}^2}}.
  \label{rhobard}
\end{eqnarray}
 For the value of $\rho/k_F^3$ defined earlier, the polarization
\begin{equation}
   P=\frac{x_{\rm max}^3-x_{\rm min}^3}{2}.
\label{bppol}
\end{equation}
This is an important relation as $P$ is a control parameter in experiments,
which can be varied over a wide range with fine control. Further, to study
the mixture of fermionic atoms with unequal masses, we define the dimensionless quantities
$\alpha=(m_1+m_2)/2\tilde m$, $\beta=(m_1-m_2)/2\tilde m$, and $\gamma=m_1m_2/\tilde m^2$.
In terms of these quantities $k_{\rm max/min}=k_F x_{\rm max/min}$, which 
are given as
\begin{equation}
   x_{\rm max/min}^2=(\alpha\hat\nu+\beta\hat\delta_\nu)
      \pm\sqrt{(\beta\hat\nu+\alpha\hat\delta_\nu)^2-
      \gamma\hat{\Delta}^{2}}.
  \label{bpxminmax}
\end{equation}
 While studying mixtures of different mass ratios, it is also convenient
to use the mass ratio $q=m_{1}/m_{2}$ as a generic parameter.
To study finite temperature effects, temperature is expressed in units of
fermi temperature $T_F,$ defined as $k_BT_F=\epsilon_F.$

%%%%%%%%%%%%%%%%%%%%%%%%%%%%%%%%%%%%%%%%%%%%%%%%%%%%%%%%%%%%%%%%%%%%%%%%%%%
%%%%            Section V: Results and discussions                    %%%%%
%%%%%%%%%%%%%%%%%%%%%%%%%%%%%%%%%%%%%%%%%%%%%%%%%%%%%%%%%%%%%%%%%%%%%%%%%%%

\section{Results and Discussions\label{results}}
Let us note that the dimensionless
parameters which describe the fermionic mixture are the dimensionless coupling $(k_Fa)^{-1}$, 
polarization $P$, temperature $T$ in units of $T_F$ and mass ratio $q$. The 
gap equation, Eq.(\ref{scaledgape}), together with the number density 
equations, that is, the average 
density given by Eq.(\ref{rhobard}) and polarization given by 
Eq.(\ref{bppol}),
are solved self consistently to obtain order 
parameter $\hat\Delta$, and chemical potentials $\hat\nu$ and $\hat\delta_\nu$. The 
thermodynamic potential, Eq. (\ref{delomg}) is then calculated using these parameters.
%%%%%%%%%%%%%%%%%%%%%%%%%%%%%%%%%%%%%%%%%%%%%%%%%%%%%%%
%%%                NEW PART                  %%%%%%%%%%
%%%%%%%%%%%%%%%%%%%%%%%%%%%%%%%%%%%%%%%%%%%%%%%%%%%%%%%
\begin{figure}
    \includegraphics[bb = 27 26 690 512,width=8cm,clip=]{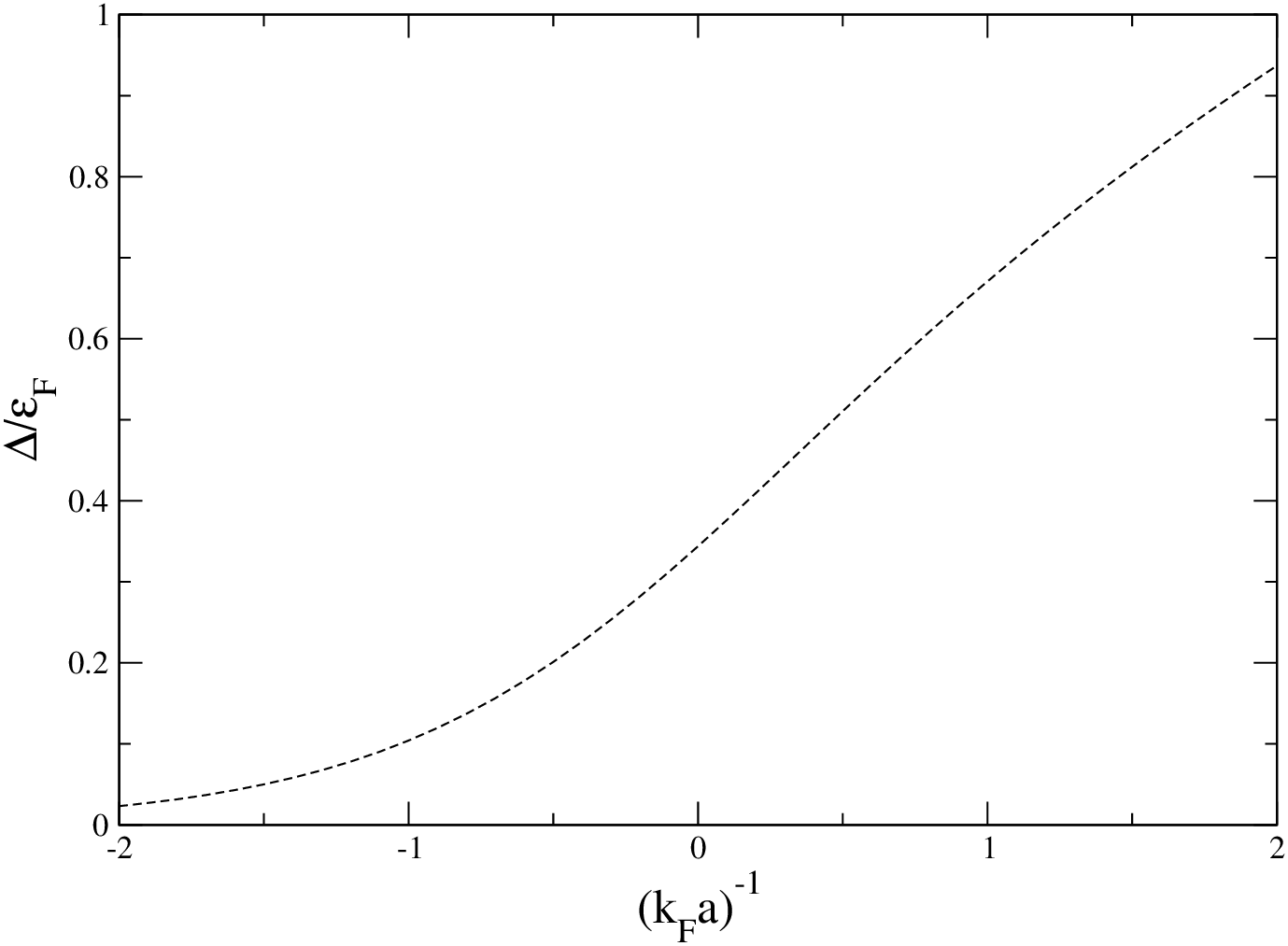}
     \caption{$\hat\Delta$ is plotted against coupling 
           $(k_Fa)^{-1}$ at zero temperature without population imbalance. 
	   %Here $\Delta$ is 
	   %expressed in units of fermi energy $\epsilon_F.$%
	   }
   \label{delta_crossover}  
\end{figure}

\begin{figure}
   \includegraphics[bb = 23 26 690 514,width=8cm,clip=]{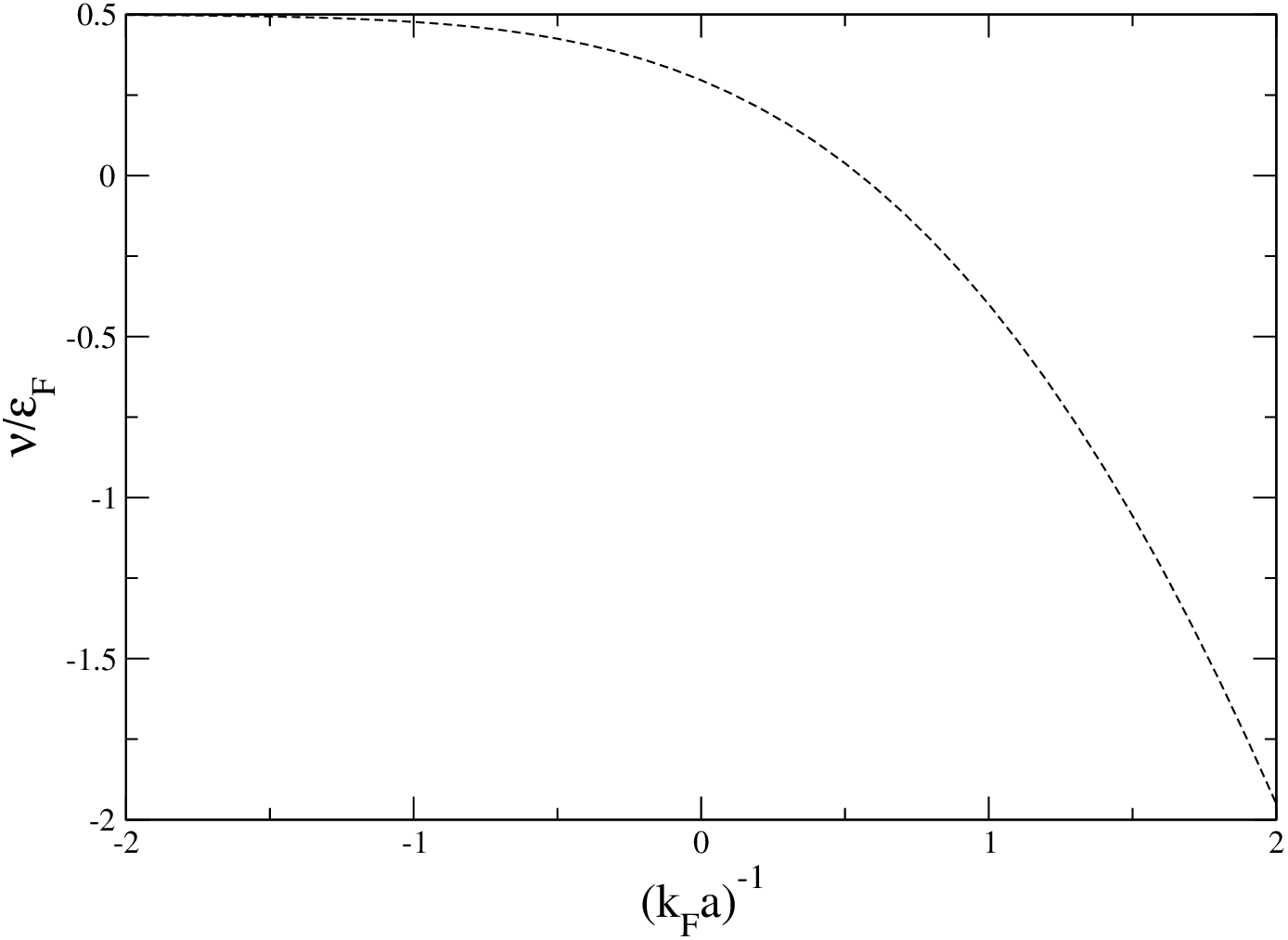}
   \caption{$\hat\nu$ is plotted against coupling 
           $(k_Fa)^{-1}$ at zero temperature without population imbalance. 
           %Here $\nu$ is expressed in units of fermi energy $\epsilon_{\rm F}.$
           }
    \label{nu_crossover}
\end{figure}

We first consider the zero polarization case. Figure \ref{delta_crossover}
shows the variation of $\Delta$ with coupling $(k_Fa)^{-1}.$ In the BCS regime
where $(k_Fa)^{-1}<-1,$  $\Delta$ is exponentially small in agreement with 
analytic solution as given in ref. \cite{stoof,amhmloff}. The variation of 
chemical potential $\nu$ is shown in the figure \ref{nu_crossover}.
The chemical potential $\nu$ decreases as the coupling $(k_Fa)^{-1}$ 
increases and it is zero at $(k_Fa)^{-1}\approx 0.55$. It is negative
at higher values of $(k_Fa)^{-1}$, which  indicates the formation
of Bose-Einstein condensation of diatomic molecules of fermionic atoms.

Next we consider the case of  nonzero polarization, however, 
for symmetric mass case i.e. with $m_1=m_2$.
In such a case, the breached pair phase shall have  fermionic gapless modes. 
The gapless modes 
occur when $\omega_i=0.$ Without loss of generality,  we shall assume here $\delta_\xi$ as negative.
For mass symmetric case, $q=1$ and hence $\delta_\xi=-\delta_\nu.$ In this case only 
$\omega_1$ can become zero.
Thus gapless mode means $\omega=|\delta_\xi|.$ 
%This can lead lead to two solution as discuused above (Eq. \ref{kminmax}).
Depending on the chemical potentials, breached pair solution 
can exist either with one ($\nu_2 <0$) or two ($\nu_1, \nu_2 >0$) fermi surfaces
referred to as BP1 and BP2 phase respectively \cite{wilczek}.

\begin{figure}[htp]
   \includegraphics[bb =37 51 686 507,width=8cm,clip=]{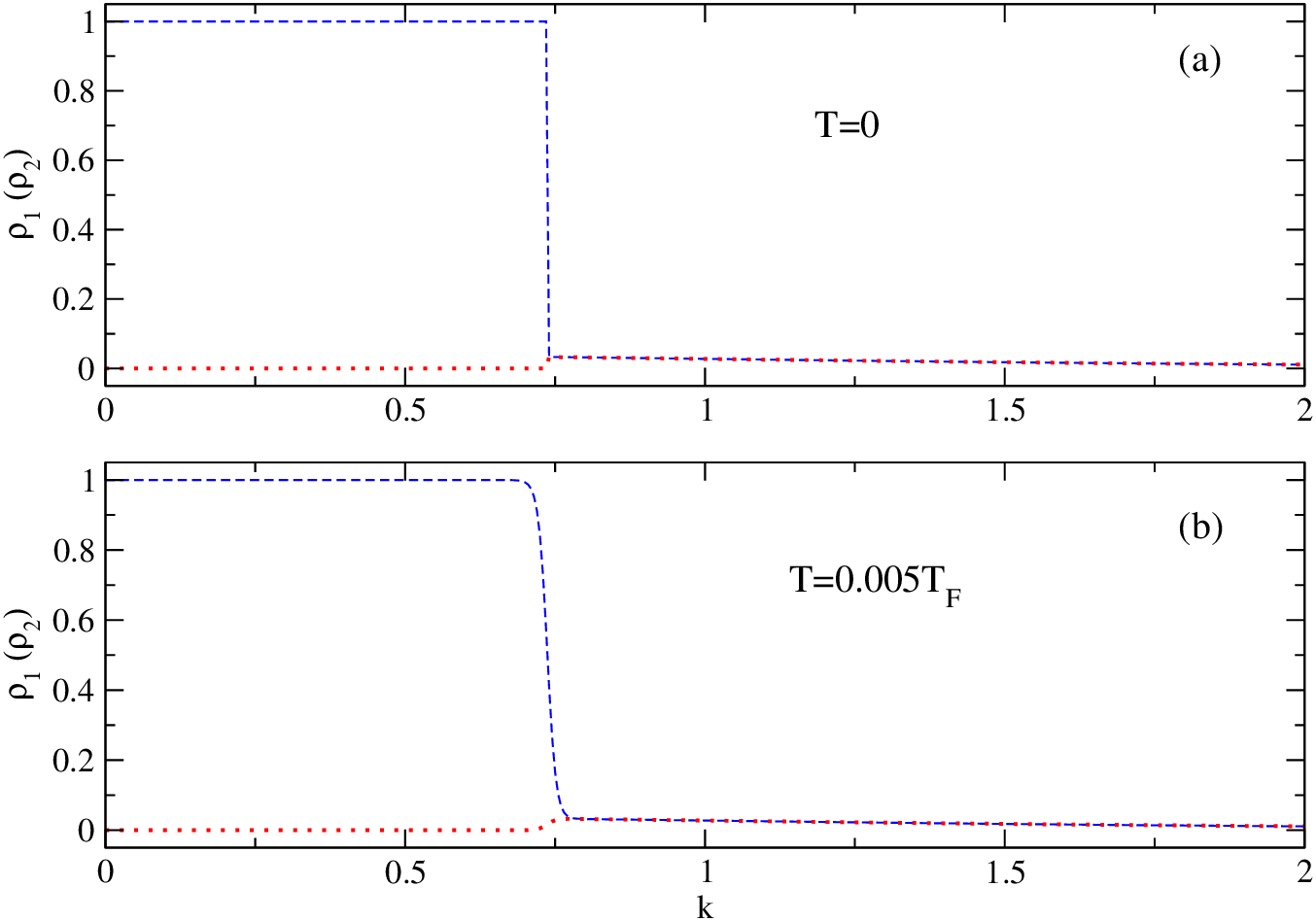}
   \caption{ The momentum space density profile for mass
            symmetric case $q=1$ at (a) $T=0$ (b) $T=0.005T_{\rm F}.$
	    Here $(k_Fa)^{-1}=2$ and $P=0.2.$ The profile corresponds to
	    BP1 phase.
	   }
   \label{q1_BP1}  
\end{figure}

The figure \ref{q1_BP1} shows the density profile for mass symmetric case at zero and 
finite temperatures in the BEC regime of interaction. The density profile here corresponds 
to $(k_Fa)^{-1}=2$ and $P=0.2$ and has the characteristic of BP1 phase of a single fermi surface.
We also verify here that the thermodynamic potential difference between the paired phase and the
normal matter is negative indicating its stability.
The critical polarization $P_c$ up to which this phase is stable for  this coupling is $P_c=0.52.$
As the coupling increases, the critical polarization $P_c$ increases and  finally reaches $P_c=1$
at $(k_fa)^{-1}\approx2.3$. The upper and lower curves correspond to zero temperature and
$T/T_F=0.005$. Finite temperature effects smoothen the distribution functions.

\begin{figure}[htp]
   \includegraphics[bb = 37 51 688 507,width=8cm,clip=]{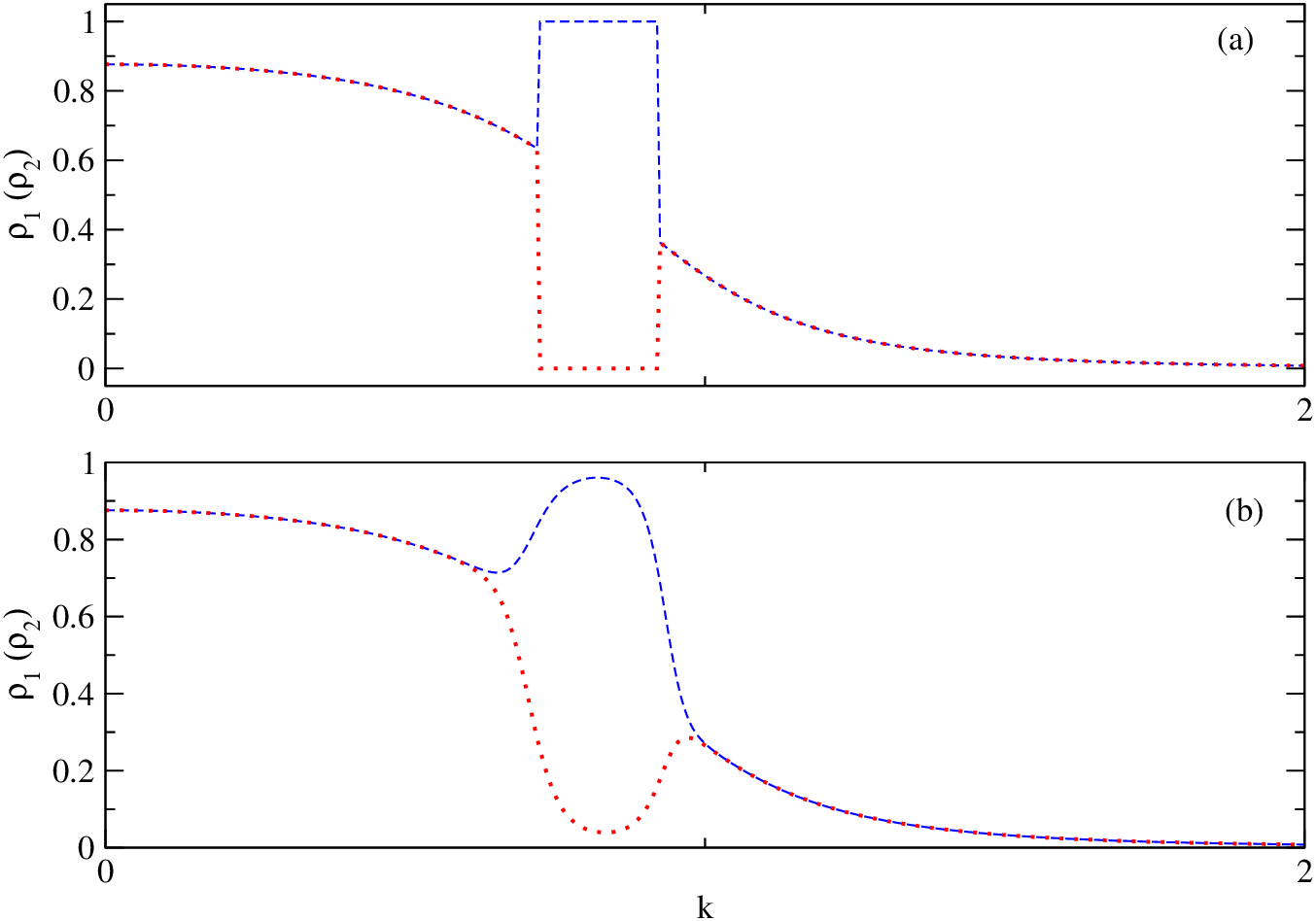}
   \caption{ The momentum space density profile for mass
            symmetric case $q=1$ at (a) $T=0$ (b) $T=0.005T_{\rm F}.$
            Here $(k_Fa)^{-1}=0$ and $P=0.2.$ The profile corresponds to
            BP2 phase. This phase, however, is unstable.
            }
   \label{q1_BP2}
\end{figure}

We also observe that, at unitarity ($(k_fa)^{-1}$=0), breached pair solution exists with
two fermi surfaces (BP2). However it is 
thermodynamically unstable i.e. $\delta\Omega$ is positive. The density profile
in momentum space is shown in the 
Fig. \ref{q1_BP2}. We have also shown the effect of temperature in the density profiles
within the present mean field calculation.  As before, the density profiles get smoothened
for finite temperatures as quasi-particle density distribution is no longer 
a $\Theta(\ldots)$ function.

\begin{figure}[htp]
   \includegraphics[bb =37 51 687 508 ,width=8cm,clip=]{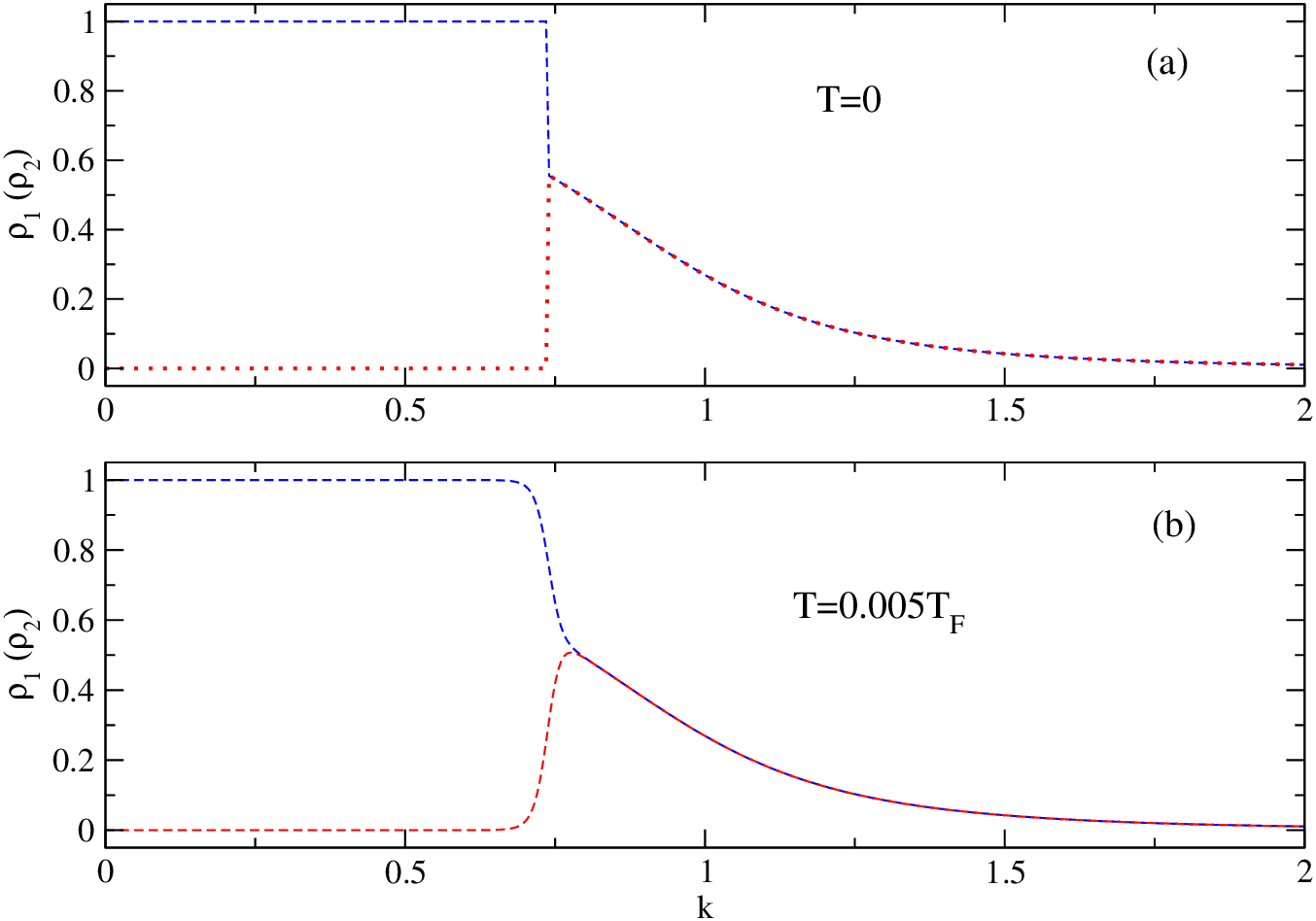}
   \caption{ The momentum space density profile for mass
            asymmetric case $q=0.15$ at (a) $T=0$ (b) $T=0.005T_{\rm F}.$
	    Here $q=m_1/m_2$, $(k_Fa)^{-1}=0.1$ and $P=0.2.$ The profile 
	    corresponds to BP2 phase. This is stable phase.
	   }
   \label{q0.15_BP1}  
\end{figure}

We next consider the mass asymmetric case with the mass ratio $q=m_1/m_2$ 
differing from unity. Specifically, we have taken
the mass ratio $q=0.15$. This ratio corresponds to $^6$Li-$^{40}$K mixture,
which has been cooled to degeneracy recently \cite{taglieber},  with $^6$Li chosen as the
majority population. The momentum space density profile for this system at 
$(k_Fa)^{-1}=0.1$ and $P=0.2.$ is shown in the Fig. \ref{q0.15_BP1}. 
Though the coupling strength
is close to unitarity, the phase corresponds to the gapless modes 
with one fermi surface (BP1).
The critical polarization turns out to be $P_c\approx 0.35.$

\begin{figure}[htp]
   \includegraphics[bb =33 42 708 523 ,width=8cm,clip=]{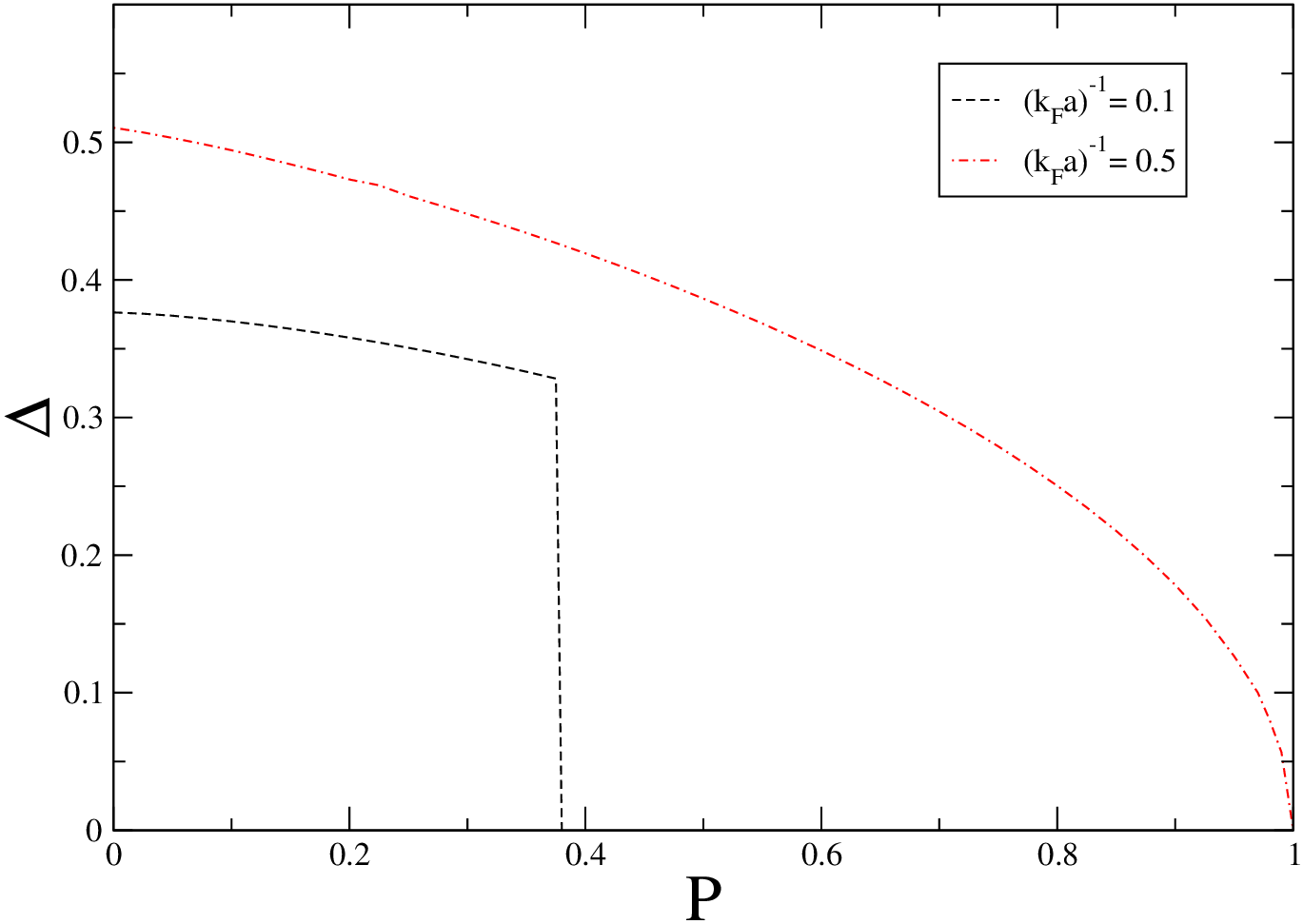}
   \caption{ The variation of $\Delta$ against polarization is shown for 
            $m_1/m_2=0.15$ at couplings $(k_Fa)^{-1}=0.1$ (dashed) and 
	    $(k_Fa)^{-1}=0.5$ (dot-dashed). 
	   }
   \label{q_pt15_p}  
\end{figure}
We see the breached pairing solution here with one fermi surface 
in the deep BEC regime. For example for  $(k_Fa)^{-1}=8$, we find
the stable BP1 state up to the critical polarization $P_c\approx 0.22.$

%%%%%%%%%%%%%%%%%%%%%%%%%%%%%%%%%%%%%%%%%%%%%%%%%%%%%%%%%%%%%%%%%%%%%%%%%%%%
%%%%%                   Section VI: Conclusion                         %%%%%
%%%%%%%%%%%%%%%%%%%%%%%%%%%%%%%%%%%%%%%%%%%%%%%%%%%%%%%%%%%%%%%%%%%%%%%%%%%
\section{summary and conclusions}\label{summary}

We have considered here a variational ground state for the system of nonrelativistic
fermions with a four fermion point interaction to model the phase structure
of the  cold atomic mixture near the Feshbach resonance. The ansatz functions
including the thermal distribution functions describing  the 
variational ground state are determined through the minimization of the
thermodynamic potential. The stability of the solutions is decided by 
comparing the thermodynamic potentials of the paired state and normal matter.

We find that gapless modes with a single fermi surface solutions are possible within
in the BEC region of the coupling constant when there is a difference of number densities of the
two species. We have in particular looked into the number density difference of the two species
differing in their masses.  There is a critical polarization for a given coupling which can sustain
pairing. Beyond this value for polarization the pairing state becomes unstable. We have not calculated
the Meissner masses \cite{kitazawa} or the number susceptibility \cite{gubankova} to discuss the
stability of different phases. Instead, we have solved the gap equations and the number density
equations self consistently and have compared the thermodynamic potentials. 
In certain regions
of couplings, we have multiple solutions of the gap equations. In such cases we have 
chosen the one which has the least value for the thermodynamic potential.

The present results  however are limited by the simplified ansatz as considered
here in terms of fermion condensates. To look into the thermal effects we have
considered rather small temperatures.
At higher temperature, particularly near the critical temperature, 
the effects of fluctuations involving
corrections arising from collective modes will play an important role, 
particularly for strong coupling. Further, considering a realistic potential
rather than the point interaction as considered here as well as the calculation 
of some of the transport properties in the different phases of
cold atomic gas, will be very interesting to investigate and
can be studied within the present framework. 
Some of these calculations
are in progress and will be reported elsewhere.

\acknowledgments
AM would like to acknowledge financial support 
from Department of Science and Technology, Government of India (project
no. SR/S2/HEP-21/2006). SAS acknowledges useful discussions with P. K. Panigrahi,
R. Rangarajan, B. Deb and J. Bhatt.

%%%%%%%%%%%%%%%%%%%%%%%%%%%%%%%%%%%%%%%%%%%%%%%%%%%%%%%%%%%%%%%%%%%%%%%%%%%
%%%%                   Section VII: Bibliography                      %%%%%
%%%%%%%%%%%%%%%%%%%%%%%%%%%%%%%%%%%%%%%%%%%%%%%%%%%%%%%%%%%%%%%%%%%%%%%%%%%
\def\regal{C.A. Regal, M. Greiner and D.S. Jin, {\PRL{92}{040403}{2004}};
M. Bartenstein {\em et al}, {\PRL{92}{120401}{2004}};
M. W. Zwierlein {\em et al}, {\PRL{92}{120403}{2004}};
J. Kianast {\em et al}, {\PRL{92}{150402}{2004}};
T. Bourdel {\em et al}, {\PRL{93}{050401}{2005}}.}
\def\amhma{Amruta Mishra and Hiranmaya Mishra,
{\PRD{69}{014014}{2004}.}}
\def\amhmb{A. Mishra and H. Mishra, {\PRD{71}{074023}{2005}.}}
\def\amhmc{A. Mishra and H. Mishra, {\PRD{74}{054024}{2006}.}}
\def\amhmq{
Amruta Mishra and Hiranmaya Mishra,
{\PRD{69}{014014}{2004}};
{\PRD{71}{074023}{2005}};
{\PRD{74}{054024}{2006}}}
\def\amhmloff{A. Mishra and H. Mishra, arXiv:cond-mat/0611058.}
\def\rapid{B. Deb, A.Mishra, H. Mishra and P. Panigrahi,
Phys. Rev. A {\bf 70},011604(R), 2004.}
\def\wilczek{W.V. Liu and F. Wilczek,{\PRL{90}{047002}{2003}},E. Gubankova,
W.V. Liu and F. Wilczek, {\PRL{91}{032001}{2003}.}}
\def\ketterle{M.W. Zweierlein, A. Schirotzek, C.H. Schunck and W. Ketterle,
Science,{\bf 311},492 (2006); G.B. Patridge, W. Li, R.I Kamar, Y. Liao
and R.G. Hulet, Science, {\bf 311}, 503 (2006).}
\def\kitazawa{M. Kitazawa, D. Rischke and I. Shovkovy, Phys Lett {\bf B637}, 367, 2006.}
\def\gubankova{E. Gubankova, A. Schmitt, F. Wilczek, Phys. Rev. {\bf B74}, 064505 (2006).}
\def\taglieber{ M.~Taglieber, A.-C.~Voigt, T. Aoki, T.W. H\"ansch, and 
                K. Dieckmann, Phys. Rev. Lett. {\bf 100}, 010401 (2008).}
\def\wille{E.~Wille et al., Phys. Rev. Lett. {\bf 100}, 053201 (2008).}
\def\gurarie{V. Gurarie, L. Radzihovsky, Ann. Phys. {\bf 322}, 2 (2007);
            Daniel E. Sheehy, Leo Radzihovsky, 
            {\em ibid} {\bf 322}, 1790 (2007).}
\def\demarco{B. DeMarco and D. S. Jin, Science {\bf 285}, 1703 (1999); 
             A. G.  Truscott et al., Science {\bf 291}, 2570 (2001)}
\def\chin{C. Chin, M. Bartenstein, A. Altmeyer, S. Riedl, S. Jochim, J.
          Hecker Denschlag, and R. Grimm, Science {\bf 305}, 1128 (2004)}
\def\zwierlein{M. W. Zwierlein, J. R. Abo-Shaeer, A. Schirotzek, C. H.
              Schunk, and W. Ketterle, Nature {\bf 435}, 1047 (2005).}
\def\tfd{H.~Umezawa, H.~Matsumoto and M.~Tachiki {\it Thermofield dynamics
and condensed states} (North Holland, Amsterdam, 1982) ;
P.A.~Henning, Phys.~Rep.253, 235 (1995).}
\def\randeria{C.A.R. Sa de Melo, M. Randeria and J.R. Engelbrecht,
{\PRL{71}{3202}{1993}}.}
\def\stoof{H. T. C. Stoof, M. Houbiers, C. A. Sackett, and R. G. Hulet,
           Phys. Rev. Lett. 76, 10 - 13 (1996)}
\def\parisha{M. M. Parish, F. M. Marchetti, A. Lamacraft, and B. D. Simons,
            Nature Phys. {\bf 3}, 124 (2007).}
\def\parishb{M. M. Parish, F. M. Marchetti, A. Lamacraft, and B. D. Simons,
             Phys. Rev. Lett. {\bf 98}, 160402 (2007)
            }
\def\sheehy{D. E. Sheehy and L. Radzihovsky, 
            Phys. Rev. Lett. 96, 060401 (2006)}
\def\shin{Shin-Tza Wu and C.-H. Pao, Phys. Rev. B {\bf 74}, 224504 (2006)}
\def\caldas{P. F. Bedaque, H. Caldas, and G. Rupak,
Phys. Rev. Lett. {\bf91}, 247002 (2003); 	
Heron Caldas,Phys. Rev. A {\bf69}, 063602 (2004).}
\def\sedrakian{	A. Sedrakian, J. Mur-Petit, A. Polls, and Muther, arXiv:cond-mat/0404577v2;
A. Sedrakian, J. Mur-Petit, A. Polls, and Muther, Phys. Rev. A {\bf72}, 013613 (2005). }
\def\nardulli{ e.g. see review R. Casalbuoni and G. Nardulli,
Rev. Mod. Phys. {\bf76}, 263 (2004)}
\def\yip{Shin-Tza Wu, C.-H. Pao, and S.-K. Yip
Phys. Rev. B {\bf74}, 224504 (2006)}

\vfil
\end{document}